\documentstyle[multicol,aps,pre,epsf]{revtex}
\begin{document}
\title{Random Walks in Logarithmic and Power-Law Potentials, Nonuniversal 
Persistence, and Vortex Dynamics in the Two-Dimensional XY Model} 
\author{A. J. Bray}
\address{Department of Physics and Astronomy, The University, Manchester,
M13 9PL, United Kingdom}

\date{\today}
\maketitle

\begin{abstract}
The Langevin equation for a particle (`random walker') moving in 
$d$-dimensional space under an attractive central force and driven by 
a Gaussian white noise is considered for the case of a power-law force, 
$F(r) \sim - r^{-\sigma}$. The `persistence  probability', $P_0(t)$, that  
the particle has not visited the origin up to time $t$, is calculated 
for a number of cases. For $\sigma > 1$, the force is asymptotically 
irrelevant (with respect to the noise), and the asymptotics of $P_0(t)$ 
are those of a free random walker. For $\sigma < 1$, the noise is 
(dangerously) irrelevant and the asymptotics of $P_0(t)$ can be extracted 
from a weak noise limit within a path-integral formalism employing the 
Onsager-Machlup functional.

The case $\sigma=1$, corresponding to a logarithmic potential, is most 
interesting because the noise is exactly marginal. In this case, $P_0(t)$ 
decays as a power-law, $P_0(t) \sim t^{-\theta}$ with an exponent $\theta$ 
that depends continuously on the ratio of the strength of the potential to 
the strength of the noise. This case, with $d=2$, is relevant to the 
annihilation dynamics of a vortex-antivortex pair in the two-dimensional 
XY model. Although the noise is multiplicative in the latter case, the 
relevant Langevin equation can be transformed to the standard form discussed 
in the first part of the paper. The mean annihilation time for a pair 
initially separated by $r$ is given by $t(r) \sim r^2\ln(r/a)$ where $a$ is 
a microscopic cut-off (the vortex core size). Implications for the 
nonequilibrium critical dynamics of the system are discussed and compared 
to numerical simulation results.   
\end{abstract}

\begin{multicols}{2}
\section{Introduction}
This paper deals with two seemingly distinct topics: persistence and 
nonequilibrium critical dynamics. We shall show that they are related 
in the context of the nonequilibrium critical dynamics of the 
two-dimensional (2D) XY model with nonconserved order parameter. The link 
is established through the study of a $d$-dimensional random walker moving 
in a logarithmic potential.

Persistence phenomena are related to first-passage problems for stochastic 
processes. Consider a stochastic process $x(t)$. The `persistence 
probability', $P_0(t)$, is the probability that $x(t)$ has not crossed some 
given level, $x_c$ (often taken to be zero), up to time $t$. The probability 
distribution, $P_1(t)$, of the first-passage time (i.e.\ the first time $t$ 
for which $x(t) = x_c$) is $P_1 = -dP_0/dt$. A familiar example is the 1D 
random walk, with Langevin equation $dx/dt = \xi(t)$, where $\xi(t)$ is a 
Gaussian white noise. For initial condition $x(0)=x_0$, the probability that 
the walker has not crossed the origin ($x=0$) up to time $t$ decays as 
$P_0(t) \sim x_0/t^{1/2}$ for $t \gg x_0^2$. The power entering this 
algebraic decay has been termed the `persistence exponent', $\theta$, 
i.e.\ $\theta = 1/2$ for the 1D random walk.

Persistence phenomena have been widely studied in recent years 
\cite{p1,p2,p3,p4,p5,p6,p7,p8,p9,OCB}. Theoretical and computational studies 
include spin systems in one \cite{p1} and higher \cite{p2} dimensions, 
diffusion fields \cite{p3}, fluctuating interfaces \cite{p4}, phase-ordering 
dynamics \cite{p5}, and reaction-diffusion systems \cite{p6}. Experimental 
studies include the coarsening dynamics of breath figures \cite{p7}, 
soap froths \cite{p8}, and twisted nematic liquid crystals \cite{p9}. 
Persistence in nonequilibrium critical phenomena has been studied 
in the context of the global order parameter, $M(t)$, (e.g.\ the total 
magnetization of a ferromagnet), regarded as a stochastic process \cite{OCB}, 
but in the present work we will address a different, and more fundamental, 
aspect of nonequilibrium critical dynamics.

If a system at its critical temperature evolves from a non-equilibrium
initial state, critical correlations develop over a length scale, 
$\xi(t)$, which increases with time. According to the standard theory 
\cite{Janssen}, $\xi \sim t^{1/z}$ for large $t$, where $z$ is the critical 
exponent for {\em equilibrium} critical dynamics. This result, which is in 
accord with simple dynamical scaling, has been demonstrated within a
field-theoretic framework \cite{Janssen}. This approach also shows that the 
result $\xi(t) \sim t^{1/z}$ is independent of the initial conditions.

In the present work we challenge this picture for the specific case of 
the 2D XY model with nonconserved order parameter, and show that 
$\xi(t) \sim t^{1/2}$ if there are no free vortices present in the 
initial state, while $\xi(t) \sim (t/\ln t)^{1/2}$ if free vortices 
are present. Physically, these two cases correspond to ordered 
initial states (e.g.\ the equilibrium state at $T=0$), and disordered 
initial states (e.g.\ the equilibrium state at $T=\infty$) respectively. 
Furthermore, since the 2D XY model is described, through the 
Kosterlitz-Thouless (KT) theory \cite{KT}, by a critical line, $T\le T_{KT}$,  
rather than a single critical point, the above dependence on initial 
conditions will persist throughout the KT phase.

The connection with persistence arises as follows. For an initial condition  
containing free vortices and antivortices, vortex-antivortex annihilation 
is the slowest relaxation process. Much can be learned by studying the 
annihilation of a single, initially widely separated, vortex-antivortex pair. 
The probability that they have not annihilated up to time $t$ defines a 
persistence problem. By a series of transformations, this can be 
mapped onto a random walk in a logarithmic potential. Analysis of this 
problem shows that the persistence exponent $\theta$ is a continuous 
function of the ratio of the strength of the potential to the strength 
of the thermal noise, i.e.\ $\theta$ is \mbox{\em nonuniversal}. 
In the context  of the vortex problem, we find $\theta = \pi\rho_s(T)/k_BT$, 
where $\rho_s(T)$ is the large-scale spin-wave stiffness at temperature $T$. 
Since the ratio $\rho_s/k_BT$ takes the universal value $2/\pi$ at $T_{KT}$, 
we obtain $\theta(T_{KT}) = 2$, while $\theta(T) \to \infty$ as $T \to 0$.

Although the vortex problem in two-dimensions was the initial motivation 
for the present study, for pedagogical reasons we will begin (section II) 
by discussing the $d$-dimensional random walk in a logarithmic potential, 
and deriving the persistence properties. We show that the exponent $\theta$ 
depends continuously on the ratio of the strength of the potential to the 
strength of the noise \cite{bouchaud}, i.e.\ the logarithmic potential is 
{\em marginal} in the Renormalization Group sense. In section III we 
consider the case of a general attractive power law force, 
${\bf F}({\bf r}) = (-A/r^\sigma) {\bf \hat{r}}$, where the logarithmic 
potential corresponds to $\sigma=1$. 
We show that for $\sigma>1$ the force is irrelevant (relative to the noise) 
as far as asymptotic persistence properties are concerned, so the results of 
the free random walk still hold. For $\sigma<1$, by contrast, the force is 
relevant and the noise becomes irrelevant. For zero noise, however, the 
dynamics are deterministic, so the persistence asymptotics are governed 
by rare fluctuations of the noise. In this sense, the noise is 
{\em dangerously} irrelevant. The asymptotic persistence follows from an 
optimal path (or steepest descent) approach formally valid in the weak 
noise limit \cite{BM,NBM}. The result is the `stretched exponential' decay, 
$P_0(t) \sim \exp[-{\rm const.}\,t^{(1-\sigma)/(1+\sigma)}]$.

Section IV deals with the application of the previous results to the 
problem of vortex-antivortex annihilation in the KT phase of the 2D XY 
model. The potential energy of a free vortex-antivortex pair is logarithmic 
in their separation, so this is the marginal case with a nonuniversal 
$\theta$. An additional complication is that the effective vortex mobility 
is scale (i.e.\ separation) dependent, implying, via the Einstein relation 
between mobility and diffusion constant, that the noise strength is also 
scale dependent or, equivalently, that the noise is multiplicative. 
However, this system can be transformed to an additive noise problem 
identical to that discussed in section II. Although the persistence exponent 
is non-universal, the dependence of the typical annihilation time on the 
initial separation $r$ has the universal form $t \sim r^2\ln(r/a)$, where $a$ 
is a (non-universal) short-distance cut-off, for all $T \le T_{KT}$. Standard 
scaling arguments then suggest that, for an initial condition containing 
many free vortices and antivortices, the characteristic length scale 
describing the approach to equilibrium will grow as 
$\xi(t) \sim (t/\ln t)^{1/2}$ throughout the KT phase.

\section{Random Walker in a Logarithmic Potential}
The Langevin equation for an isotropic, continuous-time random walker 
moving in $d$-dimensions in the central potential $V(r) = A\ln r$ is 
\begin{equation}
\frac{d{\bf r}}{dt} = -\frac{A}{r}\,{\bf \hat{r}} 
+ \mbox{\boldmath{$\xi$}}(t),
\label{Langevin}
\end{equation}
where the Gaussian white noise has the correlation function 
($i,j=1,\ldots,d$) 
\begin{equation}
\langle \xi_i(t) \xi_j(t')\rangle = 2D\,\delta_{ij}\,\delta(t-t').
\label{noise}
\end{equation}

\subsection{Reduction to a One-Dimensional Problem}
\label{sec:1d}
The analysis starts from the equivalent Fokker-Planck equation for the 
probability distribution, $P({\bf r},t)$, for the position of the particle 
at time $t$: 
\begin{equation}
\frac{\partial P}{\partial t} = \mbox{\boldmath{$\nabla$}}\cdot
\left(\frac{A}{r}{\bf \hat{r}} + D \mbox{\boldmath{$\nabla$}}P\right).
\label{FPE}
\end{equation}  
This equation can be reduced to an effectively one-dimensional equation in 
terms of the radial coordinate $r$ by integrating over the angle variables. 
Writing ${\bf r} = (r,\Omega)$, and defining the radial probability 
distribution $Q(r,t) = r^{d-1}\int d\Omega P(r,\Omega,t)$, gives 
\begin{equation}
\frac{\partial Q}{\partial t} = D \partial_r\left(\frac{b}{r}Q + 
\partial_r Q\right),
\label{FPE_r}
\end{equation}
where $\partial_r \equiv \partial/\partial r$ and 
\begin{equation}
b = \frac{A}{D} + 1-d.
\label{b}
\end{equation}
This is equivalent to the radial Langevin equation 
\begin{equation}
\frac{dr}{dt} = -\frac{A}{r} + \frac{(d-1)D}{r} + \xi(t),
\label{Langevin_r}
\end{equation}
where $\langle \xi(t)\xi(t')\rangle = 2D\,\delta(t-t')$, i.e.\ projecting 
(\ref{Langevin}) onto the radial direction leads to an additional 
repulsive force $(d-1)D/r$, proportional to the noise strength. This mean 
[see (\ref{b})] that a particle moving in a logarithmic potential in space 
dimension $d$ has the same radial distribution function as a free particle 
(random walker) moving in dimension $d' = d - A/D$ (which need not be an 
integer, or even positive).

\subsection{The Persistence Probability}
In this subsection we derive an exact expression for the persistence 
probability $P_0(t)$. First we make the change of variable 
$Q=r^{(1-b)/2}R$ in (\ref{FPE_r}).  Then the equation governing the 
relaxation modes $R(r,t) = R_k(r)\exp(-Dk^2t)$ becomes 
\begin{equation}
\frac{d^2R_k}{dr^2} + \frac{1}{r}\frac{dR_k}{dr} 
+ \left(k^2 - \frac{\nu^2}{r^2}\right)R_k = 0,
\end{equation}
where 
\begin{equation}
\nu=(1+b)/2\ . 
\end{equation}
The solutions are the Bessel functions 
$J_\nu(kr)$ and $J_{-\nu}(kr)$, so the general solution is 
\begin{eqnarray}
Q(r,t) & = & r^{(1-b)/2}\int_0^\infty dk\,[\alpha(k)J_{-\nu}(kr) 
\nonumber \\
       &  &  \hspace*{1.5cm} +\ \beta(k)J_\nu(kr)]\exp(-Dk^2t).
\label{gen-soln}
\end{eqnarray}

This desired solution has to satisfy the initial condition 
$Q(r,0)=\delta(r-r_0)$. To determine $P_0(t)$ we impose an absorbing 
boundary at $r=0$, such that the particle is removed if it reaches the 
origin. To determine the appropriate solution we note from (\ref{FPE_r}) 
that the probability current is 
\begin{equation}
j(r) = -D(\partial_r Q + bQ/r).
\label{current}
\end{equation}
Consider separately the two terms in the general solution (\ref{gen-soln}). 
The term involving  $J_{-\nu}$ behaves as $r$ for $r \to 0$, while 
the term in $J_\nu$ behaves as $r^{-b}$. Both terms have corrections which 
multiply the leading terms by power series in $r^2$. Inserting both forms 
into (\ref{current}) shows that the first term gives a finite (and negative) 
current at $r=0$, so that for this solution the origin is an absorbing 
point, or current sink. The second term, on the other hand, gives zero current 
and the origin is a not a special point: This solution 
therefore gives the spherically averaged Greens function for diffusion in 
$d'=1-b$ dimensions.

Using the orthogonality properties of Bessel functions to determine the 
functions $\alpha(k)$ and $\beta(k)$ in each case, we find that the 
Greens function is given by
\begin{eqnarray}
Q_{G}(r,t) & = & r_0\left(\frac{r}{r_0}\right)^{(1-b)/2}
\int_0^\infty dk\,k J_{-\nu}(kr_0) J_{-\nu}(kr) \nonumber \\
 & & \hspace*{3cm} \times \exp(-Dk^2t) \nonumber \\
 & = & \frac{r_0}{2Dt}\left(\frac{r}{r_0}\right)^{(1-b)/2}
 \exp\left(-\frac{r^2+r_0^2}{4Dt}\right) \nonumber \\ 
 & & \hspace*{3cm} \times I_{-\nu}\left(\frac{rr_0}{2Dt}\right),
\label{Greens}
\end{eqnarray}
where $I_\nu(z)$ is the modified Bessel function, while the relevant solution 
for an absorbing point at the origin is 
\begin{eqnarray}
Q_{abs}(r,t) & = & r_0\left(\frac{r}{r_0}\right)^{(1-b)/2}
\int_0^\infty dk\,k J_\nu(kr_0) J_\nu(kr) \nonumber \\
 & & \hspace*{3cm} \times \exp(-Dk^2t) \nonumber \\
 & = & \frac{r_0}{2Dt}\left(\frac{r}{r_0}\right)^{(1-b)/2}
 \exp\left(-\frac{r^2+r_0^2}{4Dt}\right) \nonumber \\ 
 & & \hspace*{3cm} \times I_\nu\left(\frac{rr_0}{2Dt}\right).
\label{absorbing}
\end{eqnarray}

At this point a comment on the possible values of $b$ is needed. The leading 
term in $Q_G$ for $r \to 0$ is $O(r^{-b})$, and this gives rise to zero 
current at $r=0$ as discussed above. The next-to-leading term, however, is 
$O(r^{2-b})$ leading to a current of order $r^{1-b}$. This vanishes as 
$r \to 0$ only for $b<1$. Thus the Greens function is ill defined for 
$b \ge 1$: The non-zero current at $r=0$ in this regime means the radial 
probability distribution $Q_G(r,t)$ collapses onto the origin. In terms of 
the dimension $d'$ of a free random walk we have $b=1-d'$, so the requirement 
$b < 1$ means $d'>0$, which makes physical sense. A physically reasonable 
Greens function can be restored for the case $d>0$, $A \ge dD$, where 
$b \ge 1$, by regulating the $r=0$ singularity of the force, $F(r)=-A/r$. 
For example, if $F = -A/(r + \epsilon)$ we expect the Greens function to 
have a width which vanishes with $\epsilon$.

We turn now to the more interesting case, for present purposes, of the 
distribution $Q_{abs}(r,t)$, appropriate to an absorbing point at $r=0$. 
Equation (\ref{absorbing}) gives $Q_{abs} \sim r$ for $r \to 0$, so the 
current (\ref{current}) at the origin is $j(0) = -D(1+b)(Q_{abs}/r)_{r=0}$. 
This requires $b>-1$, since $j(0)<0$ for an absorbing boundary, and $Q$ is 
necessarily non-negative. In the terms of the equivalent free random walk  
with dimension $d'$, the condition $b>-1$ requires $d'<2$, i.e.\ the 
probability to reach $r=0$ for a random walk in $d' \ge 2$ dimensions is zero.

For $b>-1$, the persistence probability, $P_0(t)$, can be readily calculated 
from (\ref{absorbing}) by first computing the current (\ref{current}) at 
$r=0$. This current gives the rate of change of the persistence, 
\begin{equation}
j(0)= \frac{dP_0}{dt} = -P_1(t)\ ,
\end{equation}
where $P_1(t)$ is the probability distribution for the time at which the 
particle first reaches the origin (`first passage time'). Using 
(\ref{current}) and (\ref{absorbing}) gives the final result:
\begin{equation}
P_1(t) = \frac{1}{\Gamma[(1+b)/2]} \frac{4D}{r_0^2} 
\left(\frac{r_0^2}{4Dt}\right)^{(3+b)/2}
\exp\left(-\frac{r_0^2}{4Dt}\right).
\label{first-passage}
\end{equation}
The persistence probability is $P_0(t) = \int_t^\infty ds\,P_1(s)$. Using 
the large-$t$ behavior of (\ref{first-passage}) gives
\begin{equation}
P_0(t) \to \frac{1}{\Gamma[(3+b)/2]}\left(\frac{r_0^2}{4Dt}\right)^{(1+b)/2}
\label{persistence}
\end{equation}
for $t \to \infty$. 
Thus the `persistence exponent', $\theta$, is given by
\begin{equation}
\theta = (1+b)/2,
\end{equation}
and is non-universal. For the free random walk in dimension $d'$, this 
translates to $\theta = (2-d')/2$.

The nonuniversality of $\theta$ with respect to the strength, $A$, of the 
potential is special to the case of a logarithmic potential, for which the 
Langevin equation (\ref{Langevin}) is invariant under the rescalings 
${\bf r} \to a {\bf r}$, $t \to a^2 t$ of space and time. 
This means that the potential is a {\em marginal} perturbation 
with respect to the equation with $A=0$, and it is this marginality which is 
responsible for the continuous variation of $\theta$ with $A$ (actually, with 
the ratio $A/D$) through it's dependence on $b$, which we recall is defined by 
(\ref{b}). The condition $b >-1$ for the particle to visit the origin with 
probability one is equivalent to $A > (d-2)D$. Note that for $d=1$ this even 
allows a (sufficiently weak) repulsive potential, whereas for $d \ge 2$ 
a guaranteed visit to the origin requires a sufficiently strong attractive 
potential.

In the following section we discuss the case of a general power law potential, 
corresponding to a force $F(r) = -A/r^{\sigma}$. We show that the force is 
irrelevant for $\sigma>1$, and the asymptotic persistence probability is that 
of a free random walker. For $\sigma <1$ the force is a relevant perturbation
to the free random walker. In this case the noise term is irrelevant, but 
`dangerously irrelevant' as far as the calculation of $P_0(t)$ is concerned. 
We show that in this case $P_0(t)$ decays as a stretched exponential.

\section{Random Walker in a Power-Law Potential}
After reducing the problem, as in the previous section, to an effectively 
one-dimensional problem for the distance, $r$, of the particle from the 
origin, the radial Langevin equation reads
\begin{equation}
\frac{dr}{dt} = -\frac{A}{r^\sigma} + \frac{(d-1)D}{r} + \xi(t),
\label{Langevin_r1}
\end{equation}

\subsection{Scaling Analysis}
Under the scale transformations $r \to a r$, $t \to a^z t$, the Langevin 
equation (\ref{Langevin_r1}) retains the same form, but with potential 
strength $A$ and noise strength $D$ rescaled to
\begin{eqnarray}
A' & = & a^{z-1-\sigma} A, \\
D' & = & a^{z-2} D.
\end{eqnarray}
For $\sigma \ne 1$ these equations have two non-trivial fixed points: \\
(i) $A > 0$, $D=0$, with $z=1+\sigma$, which is stable ($D$ scales to 
zero at large time) for $\sigma <1$, and \\
(ii) $A=0$, $D > 0$, with $z=2$, which is stable for $\sigma > 1$. 
For the special case $\sigma =1$, there is a line of fixed points with 
$z=2$ and $A/D$ arbitrary. In this last case, as we have seen, the exponent 
$\theta$ depends continuously on $A/D$.

At the second fixed point, with $\sigma > 1$, the force is irrelevant: it 
falls off too rapidly with distance to affect the asymptotic 
large-time behavior. In particular, the exponent $\theta$ is given by the 
zero-force value, $\theta = (2-d)/2$ ($0<d<2$).

At the first fixed point, with $\sigma <1$, the {\em noise} is irrelevant. 
If we set $D$ to zero, the process (\ref{Langevin_r1}) is {\em deterministic}: 
A particle staring at $r_0$ reaches the origin in a time 
$t=r_0^{1+\sigma}/[(1+\sigma)A]$. The limiting behavior of $P_0(t)$ at large 
$t$ is therefore dominated by rare events: the noise is a {\em dangerously} 
irrelevant variable in this context, and we cannot simply set it to zero. 
Instead, we have to examine the limit of small but non-zero $D$.

We can argue for the asymptotic form of $P_0(t)$ as follows. A long survival 
time of the particle is a rare event, dominated by an activated process where 
the particle initially moves {\em away from the origin}. The potential 
corresponding to the force $-A/r^\sigma$ is $V(r) = Ar^{1-\sigma}/(1-\sigma)$. 
Suppose the particle is driven (by the noise) to a point $r_1$. The time 
for a subsequent deterministic descent to the origin is 
$t = r_1^{1+\sigma}/[A(1+\sigma)]$.
The activation barrier for the `uphill' process is 
$\Delta V = [A/(1-\sigma)](r_1^{1-\sigma} -r_0^{1-\sigma})$, so the 
probability to reach $r_1$ before the origin is of order 
$\exp(-[A/(1-\sigma)]r_1^{1-\sigma}/D)$, where the term in $r_0$ in $\Delta V$ 
has been taken out and absorbed into a pre-exponential factor. Using the time 
for the subsequent free descent to estimate $r_1$, i.e.\ 
$r_1 \simeq [A(1+\sigma)t]^{1/(1+\sigma)}$, gives 
\begin{equation}
P_0(t) \sim \exp\left(-\frac{A}{(1-\sigma)D}\,
[A(1+\sigma)t]^{(1-\sigma)/(1+\sigma)}\right), 
\label{stretched}
\end{equation}
a `stretched exponential' form, $P_0(t) \sim \exp(-{\rm const.}\,t^\beta)$, 
with $\beta = (1-\sigma)/(1+\sigma)$. Note that the coefficient of $t^\beta$ 
does not depend on the initial displacement $r_0$.  The reason is that 
the rare trajectories for which the particle survives a long time take 
the particle far from its original position.

Despite its crudeness, this argument gives the correct result up to 
a constant of order unity in the exponent. This is shown in the next 
subsection using a path-integral formalism, augmented by a steepest 
descent calculation valid for $D \to 0$.  Finally we note that the 
crucial factor $(At)^{(1-\sigma)/(1+\sigma)}/D$ in the exponent of 
(\ref{stretched}), and in particular the value of the exponent $\beta$,  
can be deduced immediately from dimensional analysis once one recognizes 
that (i) the result must be independent of $r_0$, and (ii) the factor 
$1/D$ is a necessary consequence of activated dynamics.

\subsection{Path Integral Formulation for $\sigma <1$}
We begin from the probability distribution functional for the noise history,
$\xi(t)$. Since the noise is Gaussian and white, this functional is 
\begin{equation}
P[\xi(t)] = N \exp\left(-\frac{1}{4D}\int dt\,\xi^2(t)\right),
\end{equation}
where $N$ is a normalization constant. 
This can be transformed to a probability distribution functional for $r(t)$ 
using the Langevin equation (\ref{Langevin_r1}):
\begin{equation}
P[r(t)] = N\,J[r(t)]\,\exp\left(-\frac{1}{D}S[r(t)]\right)
\end{equation}
where $J[r]$ is the Jacobian of the transformation from $\xi(t)$ to $r(t)$, 
whose precise form will not concern us, and $S[r]$ is the Onsager-Machlup 
functional (or `action'),
\begin{equation}
S[r] = \frac{1}{4}\int dt\,\left(\frac{dr}{dt} + \frac{A}{r^\sigma} 
- (d-1)\frac{D}{r}\right)^2.
\label{action}
\end{equation}

It is convenient to compute $P_1(t)$, the probability density for the first 
visit to the origin (recall that $P_1 = -dP_0/dt$), given that the particle 
starts from $r_0$ at $t=0$. This is given by the path integral
\begin{equation}
P_1(t) \sim \int dr(t)\,J[r(t)]\,\exp\left(-\frac{1}{D}S[r(t)]\right),
\label{PI}
\end{equation}
where the time integral in (\ref{action}) now runs from $0$ to $t$ 
(and we introduce a dummy time integration variable $s$), and the path 
integral is over all paths $r(s)$ which satisfy $r(0)=r_0$ and $r(t)=0$. 
The use of $\sim$ in (\ref{PI}) means we are concerned only with the 
leading exponential terms and not with the prefactors.

In the limit $D \to 0$, the path integral can be evaluated by steepest 
descents. To leading order the Jacobian, $J[r(t)]$, can be replaced by
$J[r_c(t)]$, where $r_c(t)$ is the `classical' path which minimizes 
the action $S[r]$. The Jacobian therefore contributes to the prefactor, 
and we will not consider it further. In a similar way, the term $(d-1)D/r$ 
in $S$ is subdominant for $D \to 0$ and can be dropped to leading order.

As noted above, the dominant path (path of least action) is the one which 
keeps $\xi(t)$ as small as possible for as long as possible. This is achieved 
by having the particle move initially away from the origin (`uphill'), where 
a smaller noise force is needed to overcome the deterministic force driving 
the particle towards the origin. The path $r_c(t)$ therefore consists of two 
parts: an `uphill' part to a maximum displacement $r_1$, followed by a 
deterministic ($\xi=0$) downhill part. Only the uphill path has a nonzero 
action associated with it.

The variational problem for the uphill path is simplified \cite{NBM} by 
introducing the velocity $v = dr/ds$ and parametrizing the path by $v(r)$ 
instead of $r(s)$. The action then becomes
\begin{equation}
S[v] = \frac{1}{4}\int_{r_0}^{r_1}\frac{dr}{v}\,
\left(v + \frac{A}{r^\sigma}\right)^2.
\end{equation}
The variational equation $\delta S/\delta v = 0$ becomes 
$1 - A^2/(vr^\sigma)^2=0$, with solutions $v = \pm A/r^\sigma$. The minus 
sign corresponds to the deterministic downhill path, with zero action, the 
plus sign to the uphill path. The action for the latter is 
\begin{equation}
S = A\int_{r_0}^{r_1} \frac{dr}{r^\sigma} = 
\frac{A}{1-\sigma}\left(r_1^{1-\sigma} - r_0^{1-\sigma}\right). 
\label{S}
\end{equation}
The times $t_u$, $t_d$ associated with the uphill and downhill paths 
respectively, are 
\begin{eqnarray}
\label{tu}
t_u & = & \int_{r_0}^{r_1} \frac{dr}{v} 
= \frac{1}{A(1+\sigma)}[r_1^{1+\sigma} - r_0^{1+\sigma}] \\
t_d & = & \frac{r_1^{1+\sigma}}{A(1+\sigma)}.
\label{td}
\end{eqnarray}

The final step is to eliminate $r_1$ in (\ref{S}) in favor of the total 
time $t=t_u+t_d$. For $t \to \infty$, $r_1 \to \infty$ and $r_0 \ll r_1$ 
in (\ref{tu}). Dropping the terms in $r_0$ in (\ref{S}) and (\ref{tu}) 
gives $r_1 \simeq [A(1+\sigma)t/2]^{1/(1+\sigma)}$ and 
\begin{eqnarray}
P_1(t) & \sim & \exp(-S/D) \nonumber \\
& \simeq & \exp\left(-\frac{A}{(1-\sigma)D}
\left[\frac{A(1+\sigma)}{2}t\right]^{(1-\sigma)/(1+\sigma)}\right). 
\label{P1}
\end{eqnarray}
The persistence probability, $P_0(t) = \int_t^\infty dt\,P_1(t)$, clearly 
has the same asymptotic form. It differs from (\ref{stretched}) only by 
factors of order unity, as promised. It is easy to show that the corrections 
due to keeping the $r_0$ term in (\ref{tu}) vanish for $t \to \infty$, 
while the correction associated with the $r_0$ term in (\ref{S}) represents 
a time-independent prefactor. This prefactor, 
$\exp[Ar_0^{1-\sigma}/(1-\sigma)D]$ can of course be very large
(and very sensitive to $r_0$) for small $D$.

It is important to note that while the asymptotic form (\ref{P1}) was 
derived in the limit $D \to 0$, it actually holds as an asymptotic result 
for all $D$, since $D$ is an irrelevant variable and scales to zero as 
$t\to \infty$.

\section{Vortex-Antivortex Annihilation in the 2D XY Model}
\subsection{Nonequilibrium Critical Dynamics}
As a final application of these methods, we consider vortex dynamics in 
the two-dimensional (2D) XY model, which was in fact the motivation for 
the present study. At all temperatures $T \le T_{KT}$, where $T_{KT}$ is 
the Kosterlitz-Thouless (KT) transition temperature, a system prepared in a 
nonequilibrium initial state will approach the equilibrium state through 
a coarsening mechanism in which local equilibrium is established over 
a length scale $\xi(t)$ which grows with time. For example, the spin-spin  
correlation function has, according to the KT theory \cite{KT}, 
the equilibrium form $C(r) \sim r^{-\eta(T)}$, for all $T \le T_{KT}$. 
Consider now a system prepared in a random initial state, with only 
short-ranged spatial correlations, and allowed to evolve in contact 
with a heat bath at temperature $T \le T_{KT}$. According to the 
conventional theory of nonequilibrium critical phenomena \cite{Janssen}, 
the system will approach equilibrium via a dynamical scaling state, 
characterized by a growing length scale $\xi(t)$. For example, the scaling 
form for the spin-spin correlation function reads
\begin{equation}
C(r,t) = \frac{1}{r^\eta}\,f\left(\frac{r}{\xi(t)}\right).
\label{dyn_scaling}
\end{equation}
The scaling function $f(x)$ has the limiting behavior $f(0)={\rm const.}$ 
(so that equilibrium is recovered for $\xi(t) \to \infty$), while $f(x)$ 
falls off rapidly for $x \gg 1$, representing the fact that the spins are 
uncorrelated on length scales large compared to $\xi(t)$.

The standard theory of nonequilibrium critical dynamics predicts that the 
length scale $\xi(t)$ should grow as $\xi(t) \sim t^{1/z}$, where $z$ is 
the {\em equilibrium} dynamical exponent. This should hold independent 
of the initial conditions, though the scaling function $f(x)$ in 
(\ref{dyn_scaling}) can depend on initial conditions. A commonly considered 
case is uniform initial conditions (all spins parallel). For this case one 
requires that $f(x) \sim x^\eta$ for $x \to \infty$, since the long-range 
order present in the initial condition will persist at any finite time.

In a recent paper, however, Bray et al.\ \cite{BBJ} have argued, on the 
basis of numerical simulations and physical arguments, that this picture 
breaks down for the 2D XY model with nonconserved dynamics (`Model A' of 
the Hohenberg-Halperin \cite{HH} classification). Specifically, the growth 
of $\xi(t)$ will depend on whether the initial state contains free vortices and 
antivortices. In particular, they argue that for a uniform initial 
condition, for which there are no free vortices, $\xi(t)$ is determined 
by spin wave theory to be $\xi(t) \sim t^{1/2}$ \cite{RB95}. On the other hand, 
for a random initial condition there are many free vortices and antivortices 
present. The dominant coarsening mechanism in this case is vortex-antivortex 
annihilation, and this leads to $\xi(t) \sim (t/\ln t)^{1/2}$.

\subsection{Vortex-Antivortex Annihilation}
Physical arguments for $\xi(t) \sim (t/\ln t)^{1/2}$ have been given 
previously for the coarsening dynamics at $T=0$ from a random initial 
condition \cite{Yurke,RB2}. The basic idea \cite{Yurke} is to consider 
a single vortex-antivortex pair, and to derive expressions for the 
energy, $E_p(r)$ of the pair, and hence the force, $F(r)=-dE_p/dr$, 
between them, as a function of their separation $r$. The result is 
$E_p \sim \ln(r/a)$ (where $a$, the vortex core scale, is a microscopic 
length), and hence $F \sim -1/r$. Crucially it is found \cite{Yurke} that the 
vortex mobility $\mu$, which relates the force, $F$, to the velocity $v$, 
via $v=\mu F$, also depends logarithmically on the pair separation: 
$\mu \sim 1/\ln(r/a)$. At $T=0$, therefore, the variable $r$ obeys the 
deterministic equation 
\begin{equation}
\ln\left(\frac{r}{a}\right)\frac{dr}{dt} = -\frac{1}{r},
\end{equation}
up to an overall constant which can be absorbed into the timescale. 
Integrating this equation gives the annihilation time for a pair initially 
separated by a distance $r_0 \gg a$, namely $t \sim r_0^2\ln(r_0/a)$, which 
one can invert to obtain $r_0 \sim [t/\ln(t/a^2)]^{1/2}$. This already 
suggests that, in a many vortex situation, lengths and time are related by 
$\xi(t) \sim (t/\ln t)^{1/2}$.

This result can be motivated in another way using scaling arguments. If the 
characteristic scale in a many-vortex system is $\xi$, the typical force on 
a vortex scales as $1/\xi$, the typical mobility as $1/\ln(\xi/a)$ and 
the typical velocity as $d\xi/dt$. This gives $\ln(\xi/a)\,d\xi/dt \sim 
1/\xi$, i.e.\ $\xi \sim (t/\ln t)^{1/2}$ as before.

We now discuss the influence of thermal fluctuations on the annihilation 
of a single vortex-antivortex pair at non-zero temperatures.

\subsection{Vortex-Antivortex Annihilation at $T>0$}
As a first step, we present a detailed and quantitative treatment of the 
$T=0$ arguments employed in the previous subsection. It is convenient to 
adopt a continuum approach based on the non-linear sigma model Hamiltonian 
\begin{equation}
H = \frac{\rho_s}{2} \int d^2r\,(\nabla\vec{\phi})^2,
\end{equation}
where $\vec{\phi}$ is the two-component order parameter field, subject 
to a local constraint $\vec{\phi}^2=1$, and $\rho_s$ is the spin-wave 
stiffness. For a field configuration describing a single free vortex, 
$\vec{\phi} = \vec{r}/|\vec{r}|$, one has $(\nabla\vec{\phi})^2 = 1/r^2$, 
leading to an energy $E_v = (\rho_s/2)\int (d^2r/r^2) = \pi\rho_s\ln(L/a)$, 
where $L$ and $a$ are the system size and microscopic cut-off 
respectively. A vortex-antivortex pair, separated by distance $r$, screen 
each other's far fields at scales larger than $r$, leading to a pair energy 
$E_p(r) \simeq 2\pi\rho_s\ln(r/a)$, and an attractive force $F = -dE_p/dr 
= -2\pi\rho_s/r$ between the vortex and the antivortex.

The corresponding continuum description of the nonconserved (`model A') 
dynamics is given, at $T=0$, by the Langevin equation \cite{HH}
\begin{equation}
\frac{\partial\vec{\phi}}{\partial t} 
= -\Gamma \frac{\delta H}{\delta\vec{\phi}}. 
\end{equation}
This equation can be used \cite{Yurke} to compute the effective friction 
constant $\gamma(r) = 1/\mu(r)$, where $\mu$ is the mobility, associated 
with the motion of the vortex and antivortex under the force $F$ . 
An isolated vortex moving at speed $v$ in the $x$-direction has field 
configuration $\vec{\phi}(x,y,t) = \vec{\phi}_v(x-vt,y)$. Energy is 
dissipated at a rate 
\begin{eqnarray}
\frac{dE}{dt} & = & \int d^2r\,
\left(\frac{\delta H}{\delta \vec{\phi}}\right)\cdot
\left(\frac{\partial\vec{\phi}}{\partial t}\right) \nonumber \\ 
& = & -\frac{1}{\Gamma}\int d^2r\,
\left(\frac{\partial\vec{\phi}}{\partial t}\right)^2 \nonumber \\
& = & -\frac{v^2}{\Gamma}\int d^2r\, 
\left(\frac{\partial\vec{\phi}_v}{\partial x}\right)^2 \nonumber \\ 
& = & -\gamma_v v^2,  
\end{eqnarray}
where the notation $\gamma_v$ indicates the residual dependence of 
$\gamma$ on $v$ at non-infinitesimal velocities. 
Inserting the equilibrium vortex configuration, which is isotropic, gives  
the limiting zero-velocity friction constant: $\gamma_0 = E_v/\rho_s\Gamma 
= (\pi/\Gamma)\ln(L/a)$, i.e.\ $\gamma_0$, like the vortex energy 
$E_v$, diverges logarithmically with the system size, $L$. For a 
vortex-antivortex pair, this translates into a logarithmic dependence on 
the separation \cite{Yurke}, 
\begin{equation}
\gamma(r) \simeq \left(\frac{\pi}{\Gamma}\right)
\ln \left(\frac{r}{a}\right).
\end{equation}

The effect of thermal fluctuations, neglected up to now, is twofold. 
Firstly, as in the equilibrium theory, thermally activated bound 
vortex-antivortex pairs lead to a renormalization of the spin-wave 
stiffness, $\rho_s$, and kinetic coefficient, $\Gamma$, to 
temperature-dependent functions, $\rho_s(T)$ and $\Gamma(T)$, that describe 
the large length-scale properties of the system. In equilibrium, however,   
there are no {\em free} vortices at any temperature below the KT transition 
temperature, $T_{KT}$. This means that the large-scale properties are 
described by the spin-wave theory. In this theory all vortices are neglected 
and the angle representation, 
$\vec{\phi}({\bf r}) = [\cos\theta({\bf r}),\sin\theta({\bf r})]$ 
is employed, with the angles $\theta({\bf r})$ defined on the interval 
$(-\infty,\infty)$. The effective Hamiltonian for the long-wavelength 
degrees of freedom is $H = [\rho_s(T)/2]\int d^2r\,(\nabla \theta)^2$, 
and the equation of motion is 
$\partial\theta/dt = -\Gamma(T)(\delta H/\delta\theta) + \xi({\bf r},t) 
=  \Gamma(T)\rho_s(T)\,\nabla^2\theta + \xi$, where $\xi({\bf r},t)$ is 
a Gaussian white noise with correlator given by the fluctuation-dissipation 
theorem, $\langle \xi({\bf r},t)\xi({\bf r}',t')\rangle 
= 2\Gamma(T)\,k_BT \delta({\bf r}-{\bf r}')\delta(t-t')$.

Since the equation of motion is linear, it can be solved exactly. 
The dynamic exponent is $z=2$ (for all $T\le T_{KT}$). Nonequilibrium 
properties can also be evaluated exactly in the absence of free vortices
\cite{RB95}. For nonequilibrium situations where free vortices and 
antivortices are present, e.g.\ after a quench into the KT phase 
from a disordered (high-temperature) initial condition, one can argue 
as follows. In the late stages of coarsening, when the remaining free 
vortices and antivortices are widely separated, with a typical spacing 
$\xi(t)$, the spin wave theory can be used on scales much larger than 
the microscopic scale $a$ but smaller than $\xi(t)$. For example, the 
calculation of the dynamics of a single, widely-separated 
vortex-antivortex pair would proceed as at $T=0$, but using the 
temperature-dependent function $\rho_s(T)$ and $\Gamma(T)$ that incorporate 
the effect of thermal fluctuations on smaller length scales. [Strictly one 
should use a scale-dependent spin wave stiffness and kinetic coefficient, 
evaluated at scale $\xi(t)$, but we are interested in the limit 
$\xi(t) \to \infty$ where these functions can be replaced by their 
infinite-scale limits, $\rho_s(T)$ and $\Gamma(T)$].

In this approach the only modification of the $T=0$ results would be the 
replacement of $\rho_s$ and $\Gamma$, in the expression for the 
annihilation time of a vortex-antivortex pair, by their $T$-dependent 
generalizations, as follows.  Let ${\bf r}_1$ and ${\bf r}_2$ be the 
positions of the vortex and antivortex, 
and ${\bf r} = {\bf r}_2 - {\bf r}_1$ be their relative separation.
The equations of motion for ${\bf r}_1$ and ${\bf r}_2$, at $T=0$, 
read
\begin{eqnarray}
\gamma(r)\frac{d{\bf r}_1}{dt} & = & \frac{2\pi\rho_s}{r}{\bf \hat{r}}, 
\nonumber \\
\gamma(r)\frac{d{\bf r}_2}{dt} & = & -\frac{2\pi\rho_s}{r}{\bf \hat{r}},
\label{single}
\end{eqnarray}
where $\gamma(r) \simeq (\pi/\Gamma)\ln(r/a)$ is the vortex mobility.
Subtracting these gives the equation for the relative separation ${\bf r}$:
\begin{equation}
\gamma(r)\frac{d{\bf r}}{dt} = -\frac{4\pi\rho_s}{r}{\bf \hat{r}}.
\label{deterministic}
\end{equation}
Notice that the force $-4\pi\rho_s/r$ corresponds to an effective 
potential $V_{\rm eff}=2V(r)$, where $V(r)=2\pi\rho_s\ln (r/a)$ is the energy 
of a vortex-antivortex pair.

How do we generalize this equation to $T>0$? There are two distinct effects. 
Firstly, as we have discussed, thermally induced bound vortex-antivortex 
pairs renormalize the large-scale spin-wave stiffness and kinetic coefficient 
to temperature-dependent functions $\rho_s(T)$ and $\Gamma(T)$, with non-zero 
values $\rho_s(T_{KT})$ and $\Gamma(T_{KT})$ at the transition, i.e.\ the 
spin-wave theory describes the large-distance, large-time behavior at and 
below $T_{KT}$. Above $T_{KT}$, the spin wave theory breaks down due to the 
thermal nucleation of free vortices and antivortices.

If the renormalization of $\rho_s$ and $\Gamma$ were the only effects 
of thermal fluctuations, we would conclude immediately that deterministic 
annihilation of a single vortex-antivortex pair would occur, as at $T=0$, 
on a timescale $t \sim r_0^2 \ln(r_0/a)$, where $r_0$ is the initial 
separation. A second consequence of thermal noise, however, is diffusion 
of the vortex and antivortex. This means we have to add Langevin noise 
terms to (\ref{single}). We anticipate that the strength of the noise will 
be proportional to $\sqrt{\gamma(r)}$ on account of the 
fluctuation-dissipation theorem (Einstein relation). Therefore we write a 
stochastic version of (\ref{deterministic}) in the form 
\begin{equation}
\gamma(r)\frac{d{\bf r}}{dt} = -\mbox{\boldmath{$\nabla$}}U
+ \sqrt{\gamma(r)} \mbox{\boldmath{$\xi$}}(t),
\label{stochastic}
\end{equation} 
where
\begin{equation}
\langle \xi_i(t)\xi_j(t') \rangle = 4D \delta_{ij}\,\delta(t-t'),
\end{equation}
and $U(r)$ is a central potential which, it turns out, will differ by terms 
of order $D$ from the effective potential $V_{\rm eff}(r) = 4\pi\rho_s\ln r$ suggested 
by (\ref{deterministic}). We will determine $U(r)$ from the condition that 
the stationary distribution $Q(r)$ derived from (\ref{stochastic}) satisfies 
$Q(r) \propto 2\pi r\exp[-V_{\rm eff}(r)/2D]$. The strength of the noise in 
(\ref{stochastic}) is $4D$, rather than the usual $2D$ (where $D=k_BT$) 
since the noise acting on ${\bf r}$ is the sum of independent noises of 
strength $2D$ acting on ${\bf r}_1$ and ${\bf r}_2$.

The noise term in (\ref{stochastic}) is multiplicative noise in the 
Stratonovich sense. To determine the stationary distribution, and hence 
infer the form of $U(r)$, we first write (\ref{stochastic}) in the 
canonical form
\begin{equation}
\frac{d{\bf r}}{dt} = {\bf A}(r) + g(r)\,\mbox{\boldmath{$\xi$}}(t),
\label{canonical}
\end{equation}
where 
\begin{eqnarray}
\label{g1}
g & = & \frac{1}{\sqrt{\gamma}}, \\
{\bf A} & = & -\frac{1}{\gamma}\mbox{\boldmath{$\nabla$}}U
= -g^2\,\mbox{\boldmath{$\nabla$}}U.
\end{eqnarray}
The corresponding Fokker-Planck equation is 
\begin{equation}
\frac{\partial P}{\partial t} = \mbox{\boldmath{$\nabla$}}
\cdot[-{\bf A}P + 2Dg\mbox{\boldmath{$\nabla$}}(gP)].
\end{equation}
Noting that ${\bf A}({\bf r})$ is in the radial direction, i.e.\ 
${\bf A}({\bf r}) = -g^2(r)\,(\partial_rU)\hat{\bf r}$, this equation can be 
reduced to a one-dimensional equation for the angle-averaged radial 
distribution function, $Q(r,t) = r^{d-1}\int d\Omega P(r,\Omega,t)$, 
just as in section \ref{sec:1d}:
\begin{equation}
\partial_t Q = \partial_r[g^2 (\partial_rW)Q + 2Dg\partial_r(gQ)],
\label{radialFPE}
\end{equation}
where 
\begin{equation}
W(r) = U(r) - 2D(d-1)\ln r.
\label{W}
\end{equation}

Equation (\ref{radialFPE}) can be recast as the one-dimensional Langevin 
equation
\begin{equation}
\frac{dr}{dt} = -g^2(r)\,\frac{dW}{dr} + g(r)\,\xi(t),
\end{equation}
with $\langle \xi(t)\xi(t')\rangle = 4D\delta(t-t')$, which in turn can 
be written as a Langevin equation with {\em additive} noise, via the change 
of variable $dy = dr/g(r)$:
\begin{equation}
\frac{dy}{dt} = -\frac{dW}{dy} + \xi(t).
\label{additive}
\end{equation}

We can now determine the potential $U(r)$ that corresponds to a `physical' 
potential $V(r)$. Since the effective potential for the dynamics of $y$ is 
$W(y)$, the stationary distribution for $y$ is $P_y(y) \propto \exp[-W(y)/2D]$, 
implying that the radial distribution function, $Q(r) = P_y(y)(dy/dr)$,  
is given by $Q(r) \propto [g(r)]^{-1}\exp[-W(r)/2D]$ which, using (\ref{W}), 
implies $Q(r) \propto r^{d-1}\exp\{-[U(r)-2D\ln g(r)]/2D\}$. The prefactor 
$r^{d-1}$ is just the phase-space factor for the $d$-dimensional space: the 
required stationary distribution function for motion in a potential 
$V_{\rm eff}(r)$\ [=$2V(r)$] is $Q(r) \propto r^{d-1}\exp[-V_{\rm eff}(r)/2D]$. 
Equating these two results for $Q(r)$, and using $V_{\rm eff}(r)=2V(r)$, 
where $V(r)$ is the vortex-antivortex potential energy, determines $U(r)$:
\begin{equation}
U(r) = 2V(r) + 2D\ln g(r).
\end{equation}
The final equation for $y(t)$ becomes
\begin{equation}
\frac{dy}{dt} = \left(-2\frac{dV}{dr} - \frac{2D}{g}\frac{dg}{dr} 
+ \frac{2(d-1)D}{r}\right)\,\frac{dr}{dy} + \xi(t).
\label{general}
\end{equation}

The final step is to apply this (rather general) result to the 2D XY model. 
Inserting $\gamma(r) = \gamma_0\ln(r/a)$ and $V(r) = \alpha\ln(r/a)$, where 
$\gamma_0 = \pi/\Gamma$ and $\alpha = 2\pi\rho_s$, in (\ref{general}),  
using $dr/dy = g(r) = 1/\sqrt{\gamma(r)}$, and setting $d=2$, gives
\begin{equation}
\frac{dy}{dt} = \frac{2}{\sqrt{\gamma(r)}}\left(\frac{D - \alpha}{r} 
- \frac{D}{g}\,\frac{dg}{dr}\right) + \xi(t).
\label{y}
\end{equation}
Consider now the size of the second term in the large bracket. Using 
(\ref{g1}) gives
$$
-\frac{1}{g}\,\frac{dg}{dr} = \frac{1}{2}\,\frac{d\ln\gamma}{dr}  
= \frac{1}{2r\ln(r/a)} \ll \frac{1}{r}
$$
for $r \to \infty$. This means that this term can be dropped in the 
calculation of the large-distance behavior of the system. Furthermore, 
integrating the relation $dy/dr = \sqrt{\gamma(r)}$ gives, to leading 
order for large $r$, $y = r\sqrt{\gamma(r)} = r\sqrt{\gamma_0\ln(r/a)}$, 
with corrections which are again of relative order $1/\ln(r/a)$. To leading 
order, therefore, (\ref{y}) reduces to (inserting $\alpha=2\pi\rho_s$) 
\begin{equation}
\frac{dy}{dt} = \frac{2D-4\pi\rho_s}{y} + \xi(t),
\label{yfinal}
\end{equation}
with $\langle\xi(t)\xi(t')\rangle = 4D\delta(t-t')$.

Apart from a doubling of the noise strength, $D \to 2D$, associated with 
the fact that we are dealing here with a two-body problem, the result 
(\ref{yfinal}) is identical to equation (\ref{Langevin_r}) derived in 
section \ref{sec:1d} for a particle moving in a logarithmic potential, 
except that the coordinate $y$ corresponds to $r\sqrt{\gamma_0\ln(r/a)}$ in 
the present context as a result of the scale-dependent mobility.  
Vortex-antivortex annihilation can be deemed to have occurred when $r=a$, 
the vortex core size, which corresponds to $y=0$, so our previous results 
on persistence in a logarithmic potential can be applied directly to the 
persistence problem for a vortex-antivortex pair. In particular the 
parameter $b$, which controls the first-passage time distribution, $P_1(t)$, 
and persistence probability, $P_0(t)$, according to (\ref{first-passage}) 
and (\ref{persistence}) respectively, is obtained by setting $A=4\pi\rho_s$, 
$d=2$ and $D \to 2D = 2k_BT$ in (\ref{b}) to give 
\begin{equation}
b = \frac{2\pi\rho_s(T)}{k_BT} - 1.
\end{equation}

The ratio $\rho_s(T)/k_BT$ is a decreasing function of $T$, and approaches  
the universal limit $2/\pi$ for $T \to T_{KT}$. It follows that $b$ is a 
decreasing function of $T$, diverging to infinity for $T \to 0$, and 
approaching the limiting value $3$ for $T \to T_{KT}$. Recall that the 
persistence exponent is $\theta = (1+b)/2$, so $\theta$ varies continuously 
between $\theta(0)=\infty$ and $\theta(T_{KT})=2$, where the infinite 
limiting value at $T=0$ simply reflects the deterministic collapse in the 
absence of thermal noise. The persistence exponent for vortex-antivortex 
annihilation is therefore non-universal, depending continuously on $T$: 
\begin{equation}
\theta = \frac{\pi\rho_s(T)}{k_BT},
\end{equation}
for $T \le T_{KT}$. From (\ref{first-passage}) we see that the mean 
annihilation time is finite for all $T\le T_{KT}$ (since $b \ge 3$), 
and the characteristic annihilation time scales as 
$t \sim y_0^2 \sim r_0^2\ln(r_0/a)$ where $r_0$ is the initial separation. 
We conclude that thermal fluctuations do not change the fundamental 
relation between lengthscales and timescales deduced earlier from 
zero-temperature considerations, and that the coarsening growth law, 
$\xi(t) \sim (t/\ln t)^{1/2}$, holds for all $T \le T_{KT}$.

\subsection{Comparison with Simulation Data}
\label{sec:sims}
Recent simulation data \cite{BBJ} support the conclusions of the 
above analysis. Since the  method and results have already been presented 
in a short paper \cite{BBJ}, we will just briefly recall the salient 
features here. The nearest-neighbor lattice Hamiltonian 
$H = -\sum_{\langle i,j \rangle}\cos(\theta_i - \theta_j)$ was simulated on 
square lattices of size $L\times L$, for $12 \le L \le 48$. Both uniform  
(all spins parallel) and disordered (randomly orientated spins) initial 
conditions were employed. In both cases, the `time-dependent Binder 
cumulant' \cite{BHB,Zheng} 
\begin{equation}
g(L,t) = 2 - \frac{\langle[\vec{M}(t)^2]^2\rangle}
{[\langle\vec{M}(t)^2\rangle]^2}, 
\end{equation}
where $\vec{M}(t)$ is the total magnetization at time $t$, was 
measured, and a finite-size scaling, $g(L,t)= f[\xi(t)/L]$, attempted, 
where $\xi(t)$ is a characteristic length scale at time $t$.

In \cite{BBJ}, all simulations were at the KT temperature $T = 0.90$ 
\cite{Tc}. In earlier work, Luo et al.\ \cite{Zheng2} have additionally 
studied some lower temperature within the KT phase. For uniform initial
conditions, a good scaling collapse was obtained using a characteristic 
length scale $\xi(t) \sim t^{1/z}$, with an exponent $z$ that tends to 
$2$ (from slightly smaller values) for large $L$ in accordance with the 
predictions of spin-wave theory. This is entirely reasonable since the 
uniform initial conditions contain no free vortices and subsequent thermal 
fluctuations can only produce bound vortex-antivortex pairs.

For disordered initial conditions, a naive collapse using 
$\xi(t) \sim t^{1/z}$ gives unreasonably large values of $z$, in the 
range $2.3 - 2.4$, \cite{Zheng2,BBJ}, for all temperatures studied. 
Disordered initial conditions seed the system with free vortices 
and antivortices, so the analysis presented above suggests that a 
characteristic length scale $\xi(t) \sim (t/\ln t)^{1/2}$ is 
appropriate for this case. This form indeed yields an excellent data 
collapse. In practice, the form $\xi(t) = [t/\ln(t/t_0)]^{1/2}$ 
is used, with $t_0$ a fitting parameter whose value is of order unity 
\cite{BBJ}.

The data can, however, be presented in another way. The logic of 
finite-size dynamical scaling requires that the scaling variable 
can be written as $t/\tau(L)$, where $\tau(L)$ is the relaxation time 
of the finite-size system. For uniform initial conditions, the spin-wave 
theory gives $\tau(L) \sim L^2$, which is equivalent to the previous 
scaling form, but for disordered initial conditions $\tau(L)$ is given 
by the time for the last free vortex-antivortex pair, which are typically 
separated by a distance of order $L$, to annihilate, i.e.\ 
$\tau(L) \sim L^2\ln L$. A scaling collapse using $t/[L^2\ln(L/L_0)]$ 
as scaling variable is presented in Figure 1, with $L_0=1.4$ where $L_0$ 
is a short distance cut-off expected (and found) to be of order unity. 
The scaling collapse is excellent. For large enough $L$ and $t$, this 
method of plotting the data and the previous method, using 
$t/[L^2\ln(t/t_0)]$ as scaling variable, will be indistinguishable. 
For finite times and sizes, however, they differ slightly, and the new 
method gives an improved collapse.

\begin{figure}
\narrowtext
\centerline{\epsfxsize\columnwidth\epsfbox{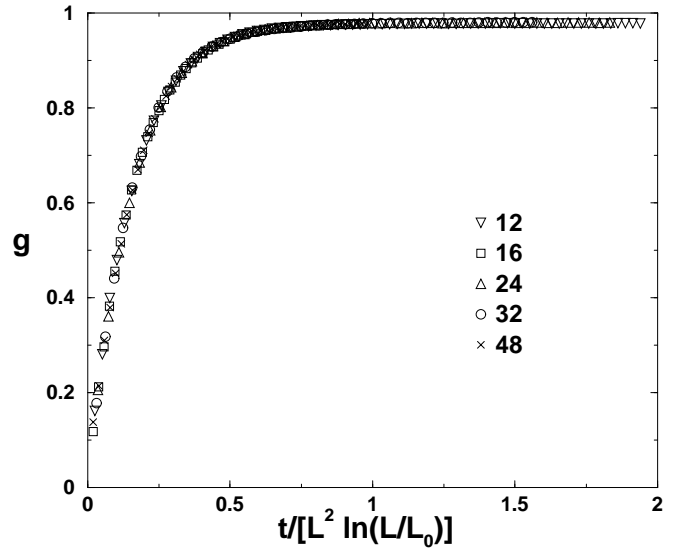}}
\caption{Scaling plot for the time-dependent Binder parameter, starting 
from a disordered initial condition, for system sizes $L=12,16,24,32,48$   
and $L_0=1.4$} 
\label{Fig.1}
\end{figure}

\subsection{Discussion}
The fact that $\xi(t)$ depends on the initial conditions is surprising 
from the viewpoint of conventional nonequilibrium critical dynamics 
\cite{Janssen}, according to which $\xi(t) \sim t^{1/z}$, where $z$ is 
the dynamical exponent for equilibrium critical dynamics. An equivalent 
statement in the context of finite-size dynamical scaling is that 
the relaxation time grows with $L$ as $\tau(L) \sim L^z$, independent  
of initial conditions. While this seems to be true for most phase 
transitions (e.g.\ the 2D Ising model \cite{Ising}) we hope we have 
convinced the reader that it does not hold for the 2D XY model, at and 
below $T_{KT}$, for disordered initial conditions. 
We conjecture that the breakdown of the standard 
field-theoretic methods \cite{Janssen} in this case is due to the 
key role played by vortex configurations which, due to their 
non-perturbative character, are not accessible to perturbative methods 
based on a $4-\epsilon$ expansion. The latter involves perturbing around 
a Gaussian theory, which cannot support topological defects. It would be 
interesting to investigate whether similar results obtain in other 
defect-driven phase transitions.

\section{Acknowledgments}
I thank Andrew Rutenberg and Alan McKane for very useful discussions, 
and Andrew Briant for refitting data from reference \cite{BBJ} in the 
form presented in Figure 1. This work was supported by the EPSRC (UK).

\end{multicols}

\end{document}